\documentclass[aps,pra,twocolumn,showpacs]{revtex4}
\usepackage{bm}
\usepackage{amsmath}
\usepackage{latexsym}
\usepackage{dcolumn}   
\newcommand{\as}[1]{\renewcommand{\arraystretch}{#1}}
\newcommand*{\centt}[1]{\multicolumn{1}{c}{#1}}

\newcolumntype{w}[1]{D{.}{.}{#1}}

\begin{document}
\preprint{Version 3.0} 

\title{Nuclear structure effects in light muonic atoms}

\author{Krzysztof Pachucki and Albert Wienczek}

\affiliation{Faculty of Physics,
             University of Warsaw,
             Pasteura 5, 02-093 Warsaw, Poland}

\date{\today}

\begin{abstract}
Nuclear structure corrections to energy levels of light muonic atoms are derived
with particular attention to the nuclear mass dependence.
The obtained result for the 2P-2S transition of $1.717(20)$~meV
serves for determination of the nuclear charge radius from the 
spectroscopic measurement in muonic deuterium. 
\end{abstract}

 \maketitle

In order to resolve discrepancies for the proton charge radius \cite{pohl, antognini, review},
spectroscopic measurements in light muonic atoms, such as $\mu$D, $\mu^3$He, and $\mu^4$He
have been performed \cite{private} for the comparison of nuclear charge radii
with those obtained from traditional atomic spectroscopy or electron 
scattering from nuclei. The nuclear charge radius can be determined from spectroscopic
measurements, provided the atomic structure is well known
and the influence of nuclear excitation on atomic levels is properly 
accounted for. The atomic structure is well understood because one can calculate 
within quantum electrodynamics the atomic levels with very high precision,
up to the value of fundamental constants. Much more problematic is
the accurate description of nuclei and their electromagnetic interactions with surrounding 
electrons and muons, because of the difficulty in solving quantum chromodynamics in the low energy scale.

The nuclear polarizability effects in muonic atoms
have been studied for some time. In 1977 Friar in \cite{friar} calculated the nonrelativistic
nuclear electric dipole polarizability and Coulomb corrections for muonic helium. 
Eighteen years later, Leidemann and Rosenfelder in \cite{rosenfelder} calculated the inelastic 
contribution for $\mu$D in a more general approach by construction of the forward
two-photon scattering amplitude for the deuteron. More recently,
we calculated in \cite{muD} nuclear structure effects in muonic deuterium
using a perturbative formalism and have shown the absence of the Zemach correction.
The results of this perturbative approach have been confirmed by Friar in \cite{friar_simp} 
using zero-range nucleon potentials.
A systematic dispersion relation approach was used in \cite{carlson2} to obtain the complete two-photon
exchange contribution, but the result suffered from insufficient inelastic
scattering data from the deuteron. Recently, 
a perturbative  approach has been pursued by independent derivation 
and numerical calculations for $\mu$He in \cite{bacca1} and $\mu$D in \cite{bacca2}. 

In this work we include higher order terms in the expansion in a small parameter
being the nuclear excitation energy over the muon mass, and recalculate
all other contributions with special emphasis on the nuclear mass dependence and  
separation of the so-called pure recoil corrections. Since the nuclear effects are the
main source of theoretical uncertainties in muonic atoms,
we aim to calculate them as accurately as possible, in order to extract
precise nuclear charge radii from the muonic atom spectroscopy. 
Our main limitation will come from the simplified model of nuclear interaction
with the electromagnetic field which assumes certain commutation relations, from  
the neglect of possible corrections to the electric dipole operator and from 
the uncertainty regarding the neutron polarizability.

In the following we derive general formulas for the nuclear 
polarizability shift using various perturbative expansions.
We aim to improve results obtained in Refs. \cite{muD, bacca1, bacca2}
by correcting mass dependencies and including higher order terms.
Let us first introduce the notation used. 
Positions of the muon and nucleons are $\vec r,\vec r_a$.
Corresponding relative positions with respect to the nuclear mass center
are $\vec\rho, \vec\rho_a$.
Momenta of the muon and nucleons are $\vec p,\vec p_a$. 
Relative nucleon momenta are $\vec q_a = \vec p_a - \vec P\,m_a/M$, where
the total nuclear momentum is $\vec P$ with the nuclear mass $M=\sum_a m_a$.
The canonical commutation relations
\begin{equation}
[r_a^i\,,\, p_b^j] = i\,\delta_{ab}\,\delta^{ij} \label{01}
\end{equation}
for relative coordinates are the following:
\begin{equation} 
[\rho_a^i\,, \,q_b^j] = i\,\biggl(\delta_{ab}-\frac{m_b}{M}\biggr)\,\delta^{ij}.\label{02} 
\end{equation}
We assume that the nuclear Hamiltonian is of the form
\begin{eqnarray}
\tilde H_N &=& \sum_a\frac{\vec p_a^{\,2}}{2\,m_a} + V_{\rm nucl} \nonumber\\
      &=& \frac{\vec P^2}{2\,M} + \sum_a\frac{\vec q_a^{\,2}}{2\,m_a} + V_{\rm nucl}\nonumber \\
      &=& \frac{\vec P^2}{2\,M} + H_N, \label{03}
\end{eqnarray}  
where $m_a$ is a proton or a neutron mass. In what follows we will neglect the isospin number,
so we will assume that each nucleon is a proton or a neutron. Under this assumption the electromagnetic 
interaction is local and is much easier to deal with. Later on, when matrix elements are calculated 
for the deuteron, the correct isospin number is assumed. This simplified treatment is because
the full description of nuclear electromagnetic interactions,
including separation of the center-of-mass motion \cite{wienczek},
have not yet been presented in the literature. 

We start derivation from the second-order Coulomb interaction in the nonrelativistic approximation
\begin{equation}
\delta E = \biggl\langle\!\phi\,\phi_N\biggl|\delta V
\frac{1}{E_N + E_0 - H_N -H_0}\,\delta V
\biggr|\phi\,\phi_N\!\biggr\rangle\,, \label{04}
\end{equation}
where
\begin{equation}
\delta V = \sum_{a=1}^Z\frac{\alpha}{|\vec \rho-\vec \rho_a|} - \frac{Z\,\alpha}{\rho}\,,
\label{05}
\end{equation}
and where $H_0$ is the nonrelativistic Hamiltonian of the muon with the reduced mass 
\begin{equation}
H_0 = \frac{\vec p^{\,2}}{2\,m_r} -\frac{Z\,\alpha}{\rho}\,. \label{06}
\end{equation}
The distance of protons from the nuclear mass center $\rho_a$ is much smaller
than that of the muon $\rho$, so the dominating contribution comes from the 
electric dipole excitations
\begin{equation}
\delta E = \alpha^2\,\biggl\langle\!\phi\,\phi_N\biggl|
\frac{\vec d\cdot\vec \rho}{\rho^3}\,
\frac{1}{E_N + E_0 - H_N -H_0}\,
\frac{\vec d\cdot\vec \rho}{\rho^3}\,
\biggr|\phi\,\phi_N\!\biggr\rangle, \label{07}
\end{equation}
where $\vec d = \sum_{a=1}^Z \vec \rho_a.$
Denoting the nuclear excitation energy by $E$,
the nonrelativistic polarizability correction is
\begin{equation}
\delta E = \frac{\alpha^2}{3}\,\int_{E_T}dE\,
|\langle\phi_N|\vec d|E\rangle|^2
\biggl\langle\phi\biggl|
\frac{\vec \rho}{\rho^3}\,\frac{1}{E_0-H_0-E}\,
\frac{\vec \rho}{\rho^3}\,\biggr|\phi\biggr\rangle
\label{08}
\end{equation}
The nuclear excitation energy $E$ is much larger than a
typical muonic atomic excitation energy,
thus one may perform the large $E$ expansion of the
muonic matrix element. The corresponding formula
for this expansion is
\begin{align}
&\biggl\langle\phi\biggl|\frac{\vec \rho}{\rho^3}\,\frac{1}{H_0-E_0+E}\,
\frac{\vec \rho}{\rho^3}\,\biggr|\phi\biggr\rangle 
= 4\,\pi\,\phi^2(0)\sqrt{\frac{2\,m_r}{E}} 
\nonumber \\ &
  + c_1\,\frac{(Z\,\alpha)^4\,m_r^4}{E} 
  + c_2\,\frac{(Z\,\alpha)^5\,m_r^4}{E}\,\sqrt{\frac{2\,m_r}{E}}
  + c_3\,\frac{(Z\,\alpha)^6\,m_r^5}{E^2},  \label{09}
\end{align}
where 
\begin{eqnarray}
\phi^2(0) &=& \frac{(m_r\,Z\,\alpha)^3}{\pi\,n^3}\,\delta_{l0},\\
c_1(2P-2S) &=& -\frac{1}{12} -\frac{1}{2}\,\ln\biggl(\frac{2\,m_r\,(Z\,\alpha)^2}{E}\biggr),\\
c_2(2P-2S) &=& \frac{19}{32} + \frac{\pi^2}{12},\\
c_3(2P-2S) &=& -\frac{7}{6}+\frac{\zeta(3)}{2}+
               \frac{5}{8}\,\ln\biggl(\frac{2\,m_r\,(Z\,\alpha)^2}{E}\biggr)
\end{eqnarray}
From this expansion,
the leading electric dipole polarizability contribution is \cite{friar}
\begin{equation}
\delta_0 E =
-\frac{4\,\pi\,\alpha^2}{3}\,\phi^2(0)\int_{E_T}dE\,
\sqrt{\frac{2\,m_r}{E}}\,|\langle\phi_N|\vec d|E\rangle|^2,
\label{13}
\end{equation}
while Coulomb corrections are
\begin{eqnarray}
\delta_{C1} E &=& \frac{Z^4\,\alpha^6\,m_r^4}{6}\!
\int_{E_T}\frac{dE}{E}
\biggl[\frac{1}{6}+\ln\biggl(\frac{2\,m_r\,(Z\,\alpha)^2}{E}\biggr)\biggr]
\nonumber \\ &&\times
|\langle\phi_N|\vec d|E\rangle|^2\,,
\label{14} \\
\delta_{C2} E &=&-\frac{Z^5\,\alpha^7m_r^3}{6}\,\biggl(\frac{19}{32}+\frac{\pi^2}{12}\biggr)\,
\int_{E_T} \!\!\!dE\biggl(\frac{2\,m_r}{E}\biggr)^{3/2}
\nonumber \\&& \times |\langle\phi_N|\vec d|E\rangle|^2\,. 
\label{15}
\end{eqnarray}
$\delta_{C3} E$ is small and thus can be neglected for light muonic atoms.
The dipole operator $\vec d$ in the above is the position
of protons with respect to the nuclear mass center.
However, one may expect some corrections to $\vec  d$. Indeed the 
chiral effective field theory predicts various relativistic corrections
to the electric dipole operator.
We do not calculate them here and therefore associate a relative uncertainty of 1\%,
which is twice the binding energy per nuclear mass.

In the evaluation of further corrections we neglect Coulomb corrections,
and so assume the on-mass-shell approximation for the muon. All corrections can therefore
be expressed in terms of the two-photon forward scattering amplitude. 
Let us consider again the related muonic matrix element $P$
for the nonrelativistic two-Coulomb exchange
\begin{equation}
P =\biggl\langle\phi\biggl|\frac{\alpha}{|\vec \rho-\vec \rho_a|}
\frac{1}{(H_0-E_0+E)} \frac{\alpha}{|\vec \rho-{\vec \rho_b}'|}\biggr|\phi\biggr\rangle.
\label{16}
\end{equation}
Using the on-mass-shell approximation and subtracting the point Coulomb exchange it becomes
\begin{eqnarray}
P &=& \alpha^2\,\phi^2(0)
\int\frac{d^3 k}{(2\,\pi)^3}\,\biggl(\frac{4\,\pi}{k^2}\biggr)^2
\biggl(E+\frac{k^2}{2\,m_r}\biggr)^{-1}
\nonumber \\ &&\times \bigl(e^{i\,\vec k\,(\vec \rho_a- \vec \rho_b')}-1\bigr).
\label{17}
\end{eqnarray}
This integral can easily be performed analytically,
but we will choose another way, which will be convenient when relativistic corrections
are included. We will calculate directly the expansion coefficients in powers of $E$.
There are two characteristic integration regions: $k\sim \sqrt{E\,m}$ and
$k\sim m$, where $m$ is the muon mass. In the first integration region, where $k$ is small, 
one performs an exponent expansion in powers of $\vec k\,(\vec \rho_a - \vec \rho_b')$. 
The leading quadratic term is the electric dipole contribution 
\begin{eqnarray}
P_0 &=& -\frac{4\,\pi}{3}\,\alpha^2\phi^2(0)\sqrt{\frac{2\,m_r}{E}}\,
\frac{(\vec \rho_a- \vec \rho_b')^2}{2}
\nonumber \\ &\rightarrow&
\frac{4\,\pi}{3}\,\alpha^2\phi^2(0)\sqrt{\frac{2\,m_r}{E}}\,\vec \rho_a\,\vec \rho_b'
\label{18}
\end{eqnarray}
and it has already been accounted for in Eq. (\ref{13}). 
The term with the fourth power of nucleon distances is
\begin{equation}
P_Q =
-\frac{2\,\pi}{15}\,m_r^2\,\alpha^2\,\phi^2(0)\,\sqrt{\frac{E}{2\,m_r}}\,
(\vec \rho_a-\vec \rho_b')^4.
\label{19}
\end{equation}  
The corresponding correction to energy is
\begin{eqnarray}
\delta_Q E &=& 
\frac{2\,\pi}{15}\,m_r^2\,\alpha^2\,\phi^2(0)
\!\int_{E_T} \!\!dE\,\sqrt{\frac{E}{2m_r}}
\nonumber \\ &&
\biggl[
\frac{10}{3}\,\langle\phi_N|\sum_a \rho_a^2|E\rangle^2
\nonumber \\ &&
-8\,\langle\phi_N|\sum_a \rho_a^i|E\rangle\,
\langle E|\sum_b \rho_b^2\,\rho_b^i|\phi_N\rangle
\nonumber \\ &&
+4\,\langle\phi_N|(\sum_a
\rho_a^i\,\rho_a^j-\delta^{ij}\,\rho_a^2/3)|E\rangle^2\biggr]
\label{20}
\\ &=& \delta_{Q0} E + \delta_{Q1} E + \delta_{Q2} E. \nonumber
\end{eqnarray}
These parts are due to the electric monopole, dipole, and the quadrupole 
nuclear excitations, correspondingly. 

In the second integration region, where $k\sim m$ is large, one performs an expansion
in powers of not exactly $E$, but of the total nuclear energy $\tilde E$, 
\begin{equation}
\tilde E = E + \frac{k^2}{2\,M}\,, \label{21}
\end{equation} 
which happens to be much more appropriate.
The first expansion term is
\begin{equation}
P = \frac{\pi}{3}\,m\,\alpha^2\,\phi^2(0)\,|\vec r_a-\vec r_b'|^3 \label{22}
\end{equation}
and the corresponding correction to energy
\begin{equation}
\delta_Z E = -\frac{\pi}{3}\,m\,\alpha^2\,\phi^2(0)\,
\sum_{a\neq b}^Z\,\langle|\vec r_a-\vec r_b|^3\rangle \label{23}
\end{equation}
is the modified Zemach moment. The second expansion term for $k\sim m$ is
\begin{equation}
P =  \alpha^2\,\phi^2(0)\,
\int\frac{d^3 k}{(2\,\pi)^3}\,\biggl(\frac{4\,\pi}{k^2}\biggr)^2\,
\biggl(\frac{2\,m}{k^2}\biggr)^2\,e^{i\,\vec k\,(\vec r_a-\vec r_b')}\,\tilde E.
\label{24}
\end{equation}
The corresponding nuclear matrix element
\begin{equation}
\langle\phi_N|
e^{i\,\vec k\cdot\vec r_a}\,(\tilde H_{N}-E_N)\,e^{-i\,\vec k\cdot\vec r_b}
|\phi_N\rangle = \frac{k^2}{2\,m_N}\,\delta_{a,b}
\label{25}
\end{equation}
is proportional to $k^2$, so the $k$ integral vanishes
after subtraction of singular terms. As a result, no corrections to
the modified Zemach moment are found. 

Consider now correction due to the finite nucleon size.
The proton and neutron charge distribution enters through the convolution 
with the Coulomb potential in Eq. (\ref{05}).
Since their charge radii are much smaller than that of nuclei, one can perform
an expansion of the electric form factors in powers of $k^2$.
When $k\sim\sqrt{E m}$, the electric dipole polarizability Eq. (\ref{13})
is modified by 
\begin{eqnarray}
P_{\rm FS} &=& \alpha^2\,\phi^2(0)
\int\frac{d^3 k}{(2\,\pi)^3}\,\biggl(\frac{4\,\pi}{k^2}\biggr)^2
\biggl(E+\frac{k^2}{2\,m_r}\biggr)^{-1}
\nonumber \\ && \times
k^2\,\frac{(r_{Ea}^2+r_{Eb}^2)}{6}\,
\frac{[\vec k\,(\vec \rho_a- \vec \rho_b')]^2}{2}
\label{26}\\ &=&
-\frac{4\,\pi}{9}\,(r_{Ea}^2+r_{Eb}^2)\,m_r^2\,\alpha^2\,\phi^2(0)\,
\sqrt{\frac{E}{2\,m_r}}(\vec\rho_a-\vec\rho'_b)^2.
\nonumber
\end{eqnarray}
The corresponding correction to energy is
\begin{eqnarray}
\delta_{\rm FS} E &=& -\frac{16\,\pi\,\alpha^2}{9}\,\phi^2(0)\,m_r^2
\nonumber \\ &&\times
\int_{E_T}\!\!dE\,\sqrt{\frac{E}{2\,m_r}}\,
\langle\phi_N|\vec d|E\rangle\,\langle E|\vec \delta|\phi_N\rangle,
\label{27}
\end{eqnarray}
where $\vec \delta = \sum_a r_{Ea}^2\, \vec \rho_a$.
When $k\sim m$ the Zemach contribution is corrected by
\begin{eqnarray}
\delta_{\rm FZ} E &=& -\frac{\pi}{3}\,m\,\alpha^2\,\phi^2(0)\,
\sum_{a}\sum_b^Z\,\frac{r_{Ea}^2}{3}\langle \vec\nabla_a^2|\vec r_a-\vec r_b|^3\rangle 
\nonumber \\
           &=& -\frac{4\,\pi}{3}\,m\,\alpha^2\,\phi^2(0)\,
\sum_{a}\sum_b^Z\,r_{Ea}^2\,\langle |\vec r_a-\vec r_b|\rangle. 
\label{28}
\end{eqnarray}
The case $a=b$ is considered separately as it involves a momentum exchange,
which is of the order of the inverse of the proton size. When a large momentum is exchanged,
the nucleon binding energy can be neglected and the muon sees free nucleons. 
The individual Zemach radii and nucleon polarizabilities are combined together
into effective Dirac-delta type interactions and are accounted for in $\delta_{N,P}E$ 
in Eqs. (\ref{45}), and (\ref{46}).

Consider now corrections from the two-Coulomb exchange
using the relativistic (Dirac) Hamiltonian for the muon. 
Equation (\ref{17}) is replaced by
\begin{eqnarray}
P &=& \alpha^2\,\phi^2(0)\,
\int\frac{d^3 k}{(2\,\pi)^3}\,\biggl(\frac{4\,\pi}{k^2}\biggr)^2\,
e^{i\,\vec k\cdot(\vec r_a-\vec r_b')}
\label{29} \\ && \nonumber
\biggl(\frac{E_k+m}{2\,E_k}\,\frac{1}{\tilde E+E_k-m} +
\frac{m-E_k}{2\,E_k}\,\frac{1}{\tilde E+E_k+m}\biggr),
\end{eqnarray}
where $E_k = \sqrt{k^2+m^2}$.
When $k\sim\sqrt{2\,m\,E}$ one employs a small $k$ expansion.
The leading term coincides with Eq. (\ref{17}). The next term is
\begin{equation}
P = -\alpha^2\,\phi^2(0)\,
\int\frac{d^3 k}{(2\,\pi)^3}\,\biggl(\frac{4\,\pi}{k^2}\biggr)^2\,
e^{i\,\vec k\cdot(\vec r_a-\vec r_b')}
\frac{\tilde E\,k^2}{(2\,m\,\tilde E + k^2)^2}
\label{30}
\end{equation}
Only the quadratic term in nuclear distances contributes,
and after subtraction of large $k$ singularities the corresponding correction to energy
\begin{eqnarray}
\delta_R E &=& \frac{2\,\pi\,\alpha^2}{3}\,\phi^2(0)
\int_{E_T}\!\!dE\,\sqrt{\frac{E}{2\,m_r}}\,|\langle\phi_N|\vec d|E\rangle|^2
\nonumber \\&&
\times \biggl(1-5\,\frac{m}{M}\biggr)
+O\Bigl(\frac{m}{M}\Bigr)^2
\label{31}
\end{eqnarray}
is in agreement with the former result of Ref. \cite{muD} in the infinite nuclear mass limit.

When $k\sim m$ one can perform the Taylor expansion of the integrand of Eq. (\ref{29})
in powers of $\tilde E$. The term without $\tilde E$, 
\begin{equation}
P = \alpha^2\,\phi^2(0)\,
\int\frac{d^3 k}{(2\,\pi)^3}\,\biggl(\frac{4\,\pi}{k^2}\biggr)^2\,
e^{i\,\vec k\cdot(\vec r_a-\vec r_b')}\,\frac{2\,m}{k^2},
\label{32}
\end{equation}
is exactly the same as in the nonrelativistic limit,
and has already been accounted for. The linear in $\tilde E$ term in Eq. (\ref{29}) is
\begin{eqnarray}
P &=& -\alpha^2\,\phi^2(0)\,
\int\frac{d^3 k}{(2\,\pi)^3}\,\biggl(\frac{4\,\pi}{k^2}\biggr)^2\, 
e^{i\,\vec k\cdot(\vec r_a-\vec r_b')}\,\tilde E
\nonumber \\ &&  \times\frac{m\,(4\,m^2 + 3 k^2)}{E_k\,k^4}. 
\label{33}
\end{eqnarray}
The corresponding nuclear matrix element can be transformed
using Eq. (\ref{25}), so correction to energy becomes
\begin{equation}
\delta'_C E = \sum_a\alpha^2\,\phi^2(0)\,
\int\frac{d^3 k}{(2\,\pi)^3}\,\biggl(\frac{4\,\pi}{k^2}\biggr)^2\, 
\frac{m\,(4\,m^2 + 3 k^2)}{2\,m_N\,E_k\,k^2} 
\label{34}
\end{equation} 
a recoil correction for each individual nucleon.
So, for $ k\sim m$ the muon sees individual nucleons
and these corrections become the sum of nucleon recoil corrections
\begin{eqnarray}
\delta'_C E &=&  -\frac{4}{3\,m}\,\alpha^2\,\phi^2(0)\,
\biggl(\frac{Z}{m_N} - \frac{Z^2}{M}\biggr)
\label{35}
\end{eqnarray}
with subtracted muon-nucleus recoil correction to avoid the double counting
with the so-called pure recoil correction. This is because
recoil corrections are by definition included in the Lamb shift
as a QED correction for a point nucleus.

The Coulomb exchange is not a complete correction; there are single and double transverse
photon exchange corrections, and their calculation is more complicated. The main reason for this
is the overlap of the nuclear recoil and nuclear polarizability corrections.
Let us repeat now the calculation by replacing the two-Coulomb exchange amplitude, Eq. (\ref{29}),
by a complete two-photon exchange.  

When $k\sim E$ or $k\sim\sqrt{E\,m}$ one can use a dipole approximation, where the coupling
of the nucleus to the electromagnetic field is $-\vec d\,\cdot\,\vec E(\vec R)$, as in Eq. (\ref{06}).
Correction to energy due to two-photon exchange in the dipole approximation is \cite{erickson}
\begin{eqnarray}
\delta E &=&
-e^4\,\phi^2(0)\,\frac{1}{3}\,\int_{E_T} dE \langle\phi_D|\vec d|E\rangle^2\,
\int\frac{d\,\omega}{2\,\pi\,i}\,
\nonumber \\ &&
\int^\epsilon\frac{d^3k}{(2\,\pi)^3}\,\frac{1}{E+\omega + k^2/(2\,M)}
\biggl(1+\frac{2\,\omega^4}{(\omega^2-k^2)^2}\biggr)\nonumber \\ && \times
\frac{4}{(\omega^2+2\,m\,\omega-k^2)(\omega^2-2\,m\,\omega-k^2)}.
\label{36}
\end{eqnarray}
The leading nonrelativistic term agrees with that in Eq.~(\ref{13}),
while the leading relativistic correction agrees with Eq.~(\ref{31}). 
The higher order correction (in powers of $E/m$) is 
\begin{eqnarray}
\delta'_R E &=& 
\frac{4}{3}\,\alpha^2\,\phi^2(0)\,\langle\phi_N|\vec d\;\frac{(H_N-E_N)}{m}
\nonumber \\ &&
\times\biggl[1+\ln\frac{(H_N-E_N)}{\epsilon}\biggr]\vec d\,|\phi_N\rangle.
\label{37}
\end{eqnarray}
Using the commutation relations of Eq. (\ref{02}) the nuclear matrix element is
\begin{equation}
\frac{4}{3}\langle\phi_N|\vec d\;(H_N-E_N)\,\vec d\,|\phi_N\rangle = 
2\,\biggl(\frac{Z}{m_N}-\frac{Z^2}{M}\biggr)\,,
\label{38}
\end{equation}
and this correction can be rewritten in the form
\begin{equation}
\delta'_R E =
\frac{2}{m}\,\alpha^2\,\phi^2(0)\,\biggl(1+\ln\frac{2\,\bar E}{m}+\ln\frac{m}{2\,\epsilon}\biggr)\,
\biggl(\frac{Z}{m_N}-\frac{Z^2}{M}\biggr)\,,
\label{39}
\end{equation}
where
\begin{equation}
\ln\bar E = \frac{\langle\phi_N|\vec d\,(H_N-E_N)\,\ln(H_N-E_N)\vec d\,|\phi_N\rangle}
                 {\langle\phi_N|\vec d\,(H_N-E_N)\vec d\,|\phi_N\rangle}\,.
\label{40}
\end{equation}
The identity (\ref{38}) is an approximate one. The electric dipole operator does not commute
with the nuclear potential. Neglected terms can be interpreted as if due to $\vec A^2$ vertex with pions.

When $k\sim m$ the complete two-photon exchange 
is a recoil correction from individual protons [see Eq. (32) of Ref. \cite{helamb}], 
\begin{eqnarray}
\delta_H E &=& \frac{4\,\pi^2\,\alpha^2}{m\,m_N}\,\phi^2(0)\,
\int_\epsilon\frac{d^3k}{(2\,\pi)^3}
\nonumber \\ &&
\times \Bigl[\frac{k^4+6\,k^2\,m^2+8\,m^4}{k^6\,\sqrt{k^2+m^2}}-\frac{1}{k^3}\Bigr]\,.
\label{41}
\end{eqnarray}

The contribution beyond the previously considered
Coulomb part, Eq. (\ref{35}), is 
\begin{equation}
\delta'_HE = \frac{2}{m}\,\alpha^2\,\phi^2(0)\,\ln\frac{2\,\epsilon}{m}\,
\biggl(\frac{Z}{m_N}-\frac{Z^2}{M}\biggr),
\label{42}
\end{equation}
where we again subtract the corresponding nuclear recoil correction.
The $\ln\epsilon$ dependence cancels out with that in Eq. (\ref{39}), as it should,
and the sum of higher order corrections is
\begin{eqnarray}
\delta_{\rm HO} E &=& \delta'_C E +\delta'_R E + \delta'_H E\nonumber\\&=&
\frac{2}{m}\,\alpha^2\,\phi^2(0)\,\biggl(\frac{1}{3}+\ln\frac{2\,\bar E}{m}\biggr)\,
\biggl(\frac{Z}{m_N}-\frac{Z^2}{M}\biggr). \nonumber \\
\label{43}
\end{eqnarray} 
These are all nuclear structure corrections up to the order $\alpha^5\,m^2/M$. 
In some cases, such as for the deuteron nucleus, where the magnetic moment is relatively large,
higher order effects due to the second-order magnetic interaction $\alpha^5 m^3/M^2$ 
may play a role. The corresponding correction for the deuteron was obtained in Ref. \cite{muD},
\begin{eqnarray}
\delta_M E &=&
\frac{8\,\,\pi\,\alpha^2}{3}\,\phi^2(0)
\biggl(\frac{g_p-g_n}{4\,m_p}\biggr)^2
\nonumber \\ && \times
\int_{E_T}\!\!\!\!dE\,\sqrt{\frac{E}{2\,m_r}}\,
\langle\phi_N|\vec s_p-\vec s_n|E\rangle^2,
\label{44}
\end{eqnarray}
but the numerical value, as was pointed out in \cite{bacca2}, was in error, so we correct it here
and present our updated value in Table I.

There are, in addition, contributions due to intrinsic elastic and inelastic two-photon exchanges 
with individual nucleons, which include the third Zemach moment and the nucleon polarizability.
While the contribution from the proton is well known from studies on muonic hydrogen \cite{carlson1},
$\Delta E(2S) = -36.9(2.4)\,\mu$eV,
less is known about the contribution from the neutron. Following \cite{carlson2},
we assume that this contribution is as large as the inelastic part for the proton 
$13.5\,\mu$eV, and associate 50\% uncertainty. Therefore, the contribution
from intrinsic nucleon polarizabilities and elastic two-photon exchanges is
\begin{eqnarray}
\delta_{P} E &=& -\frac{8\,Z^4}{n^3}\,\,\delta_{l0}\,\frac{m_{rN}^3}{m_{rH}^3}\,36.9\,{\rm meV},
\label{45}\\
\delta_{N} E &=& -\frac{8\,(A-Z)\,Z^3}{n^3}\,\,\delta_{l0}\,\frac{m_{rN}^3}{m_{rH}^3}\,13.5\,{\rm meV}.
\label{46}
\end{eqnarray}

The final expression for the nuclear polarizability combined with
the elastic contribution but with subtracted nuclear recoil of order $\alpha^5 m^2/M$ is
\begin{eqnarray}
\Delta E &=& \delta_0 E + \delta_C E + \delta_R E + \delta_Q E + \delta_{\rm FS} E + 
\delta_{\rm FZ} E 
\nonumber \\ &&
+ \delta_M E + \delta_P E +\delta_N E + \delta_{\rm HO} E + \delta_Z E.
\label{47}
\end{eqnarray}
and the elastic contribution for the neutron using Galster parametrization \cite{galster}
is found to be negligible.

Numerical results for muonic deuterium are obtained by using the 
AV18 potential \cite{av18} with the help of a discrete variable representation 
\cite{dvr} method for solving the Schr\"odinger equation,
and are presented in Table I.
\begin{table}[!htb]
\caption{Nuclear structure corrections in muonic deuterium
for 2P-2S transition. Fundamental physical constants are from Ref. \cite{NIST},
and $r_p^2= 0.8409^2$ fm$^2$, $r_n^2 = -0.1161$ fm$^2$.
$\delta^{(0)}_C$ from \cite{bacca2} includes only the logarithmic part of $\delta_{C1}E$,
which we find here to be a good approximation.}
\label{table1}
\begin{ruledtabular}\as{1.25}
 \begin{tabular}{cw{2.6}crw{2.5}}
\centt{Correction}&   \centt{Value in meV} & Eq. & Ref. \cite{bacca2} & \centt{\cite{bacca2}-AV18}\\ 
\hline
$\delta_0 E$      &  1.910   &  (\ref{13}) & $\delta^{(0)}_{D1}$         & 1.907\\
$\delta_{C1} E$    & -0.255   & (\ref{14}) & $\rightarrow\delta^{(0)}_C$ &-0.262\\
$\delta_{C2} E$    & -0.006   & (\ref{15}) & $\rightarrow\delta^{(0)}_C$ \\
$\delta_{Q0} E$    & -0.042   &  (\ref{20}) & $\delta^{(2)}_{R2}$        & -0.042\\
$\delta_{Q1} E$    &  0.139   &  (\ref{20}) & $\delta^{(2)}_{D1D3}$       & 0.139 \\
$\delta_{Q2} E$    & -0.061   &  (\ref{20}) & $\delta^{(2)}_Q$           &-0.061\\
$\delta_{\rm FS} E$ &  0.020   &  (\ref{27}) & $\delta^{(2)}_{NS}$        & 0.015\\
$\delta_{\rm FZ} E$ & -0.018   &  (\ref{28}) & $\delta^{(1)}_{np}$       &-0.017\\
$\delta_R E$      & -0.026    &  (\ref{31}) & $\rightarrow\delta^{(0)}_L+\delta^{(0)}_T$& -0.017\\
$\delta_{\rm HO} E$&  0.004    &  (\ref{43}) & $\rightarrow\delta^{(0)}_L+\delta^{(0)}_T$\\
$\delta_M E$      & -0.008   &  (\ref{44}) & $\delta^{(0)}_M$           &-0.008\\
$\delta_P E$      &  0.043(3) &  (\ref{45}) &                          & 0.0135\\
$\delta_N E$      &  0.016(8) &  (\ref{46}) &                          & 0.0135\\[1ex]
$\Delta E $       & 1.717(20) &             &                          & 1.681(20)
\end{tabular}
 \end{ruledtabular}
\end{table}
They are generally in agreement with our previous calculations \cite{muD}
with few exceptions. Differences are due to improved mass dependence
of corrections beyond the nonrelativistic dipole term. We also corrected 
the magnetic contribution, which previously was in error, and included higher order
correction $\delta_{\rm HO} E$, and most importantly the polarizability of the neutron.
We have also included, following Refs. \cite{bacca1, bacca2} finite size corrections 
$\delta_{\rm FS}$ and $\delta_{\rm FZ}$, although our results are slightly different.

In comparison to Refs \cite{bacca1, bacca2}, we agree with their numerics,
agree with the use of reduced mass in $\delta_Q E$, 
but disagree with their mass dependence of all other higher order corrections. 
Moreover, our formula for the finite size correction $\delta_{\rm FS}$ slightly disagree with 
the corresponding $\delta^{(2)}_{NS}$ due to the opposite sign for the neutron radius contribution.
Our $\delta_R E$ differs from the corresponding $\delta^{(0)}_{LT}$,
apart from mass dependence, due to the fact that for the large momentum exchange $k\sim m$,
the dipole approximation does not hold and we account for this in $\delta_{\rm HO} E$.
Finally, we consider the elastic contribution of the proton structure correction to be
a part of the overall nuclear structure correction, in contrast to Refs. \cite{carlson2, bacca2}.
Our argument is the following. The nuclear structure contribution
to high extent is given by the forward two-photon scattering of the nucleus. 
When momentum exchange is much larger than the nuclear binding energy,
muon sees individual nucleons and the total scattering amplitude is
a coherent sum of total scattering amplitudes from each nucleon.
By total we mean the elastic and inelastic contributions,
since this division is a pure convention. Therefore, both contributions should be 
included in the calculation of the Lamb shift, 
and we include them, for convenience, in the part called the nuclear structure correction.

Considering the uncertainty related to numerical evaluation of matrix elements,
Ref. \cite{bacca2} has performed calculations with AV18 potentials and with various orders of 
chiral effective field theory, finding $0.6\%$ dependence on the potential used.
Our numerical values, when the same formulas are used, are in perfect agreement
with those of Ref. \cite{bacca2}. Since we neglect possible corrections to the electric
dipole moment, which in fact depends on the model potential, we do not associate additional 
uncertainty beyond that assumed for the electric dipole polarizability. 
Regarding uncalculated higher order terms, the most significant seems to be
the Coulomb correction to $\delta_Q E$ in Eq. (\ref{20}), which we estimate by about 0.005 meV.
Therefore, the final uncertainty is determined by 50\% of the estimated neutron polarizability,
1\% of $\Delta E$ due to neglect of corrections to the electric dipole moment, and 
0.005 meV due to neglected higher order terms.

Our final result for the nuclear structure correction $\Delta E$
is not in perfect agreement with that of Ref. \cite{bacca2}, as explained above, 
mostly due to inclusion of the proton elastic contribution. 
In spite of other small discrepancies with \cite{bacca2}, the presented perturbative approach
seems to be more efficient than the dispersion relation approach of Ref. \cite{carlson2}.
If further improvements are required, the best way is probably by
joining within the dispersion relation approach, the inelastic scattering data at high energies
with nuclear model calculations at low energies, to account properly for the high energy structure 
of the deuteron.

\begin{acknowledgments}
The authors would like to acknowledge the support of the Polish National Science Center (NCN) under Grant No. 2012/04/A/ST2/00105.
\end{acknowledgments}


\begin{thebibliography}{99}
\bibitem{pohl} R. Pohl {\em et al.}, Nature (London) {\bf 466}, 213 (2010).
\bibitem{antognini} A. Antognini {\em et al}, Science {\bf 339}, 417 (2013).
\bibitem{review} R. Pohl, R. Gilman, G.A. Miller, and K. Pachucki, Annu. Rev. Nucl. Part. Sci. {\bf 63}, 175 (2013).
\bibitem{private} R. Pohl (private communication).
\bibitem{friar} J.L. Friar, Phys. Rev. C {\bf 16}, 1540 (1977).
\bibitem{rosenfelder} W. Leidemann and R. Rosenfelder, Phys. Rev. C {\bf 51}, 427 (1995).
\bibitem{muD} K. Pachucki, Phys. Rev. Lett. {\bf 106}, 193007 (2011). 
\bibitem{friar_simp} J.L. Friar, Phys. Rev. C {\bf 88}, 034003 (2013). 
\bibitem{carlson2} C. E. Carlson, M. Gorchtein, and M. Vanderhaeghen, Phys. Rev. A {\bf 89}, 022504 (2014). 
\bibitem{bacca1} C. Ji, N. Nevo Dinur, S. Bacca, and N. Barnea, Phys. Rev. Lett. {\bf 111}, 143402 (2013);
                 N. Nevo Dinur, N. Barnea, C. Ji, and S. Bacca, Phys. Rev. C 89, 064317 (2014);
                 C. Ji, N. Nevo Dinur, S. Bacca, and N. Barnea, Few-Body Syst. {\bf 55}, 917 (2014).
\bibitem{bacca2} O.J. Hernandez, C. Ji, S. Bacca, N. Nevo Dinur, and N. Barnea, Phys. Lett. B {\bf 736}, 344 (2014).
\bibitem{wienczek} A. Wienczek, M. Puchalski, and K. Pachucki, Phys. Rev. A {\bf 90}, 022508 (2014).
\bibitem{erickson} J. Bernab\'eu and T.E.O. Ericson, Z. Phys. A {\bf 309}, 213 (1983).
\bibitem{helamb} K. Pachucki, J. Phys. B {\bf 31}, 5123 (1998). 
\bibitem{carlson1} C. E. Carlson and M. Vanderhaeghen, Phys. Rev. A {\bf 84}, 020102(R) (2011). 
\bibitem{galster} T.R. Gentile and C.B. Crawford, Phys. Rev. C {\bf 83}, 055203 (2011).
\bibitem{av18} R.B. Wiringa, V.G.J. Stoks, and R. Schiavilla, Phys. Rev. C {\bf 51}, 38 (1995).
\bibitem{dvr} D. T. Colbert and W. H. Miller, J. Chem. Phys. {\bf 96}, 1982 (1992).
\bibitem{NIST} Peter J. Mohr, Barry N. Taylor, and David B. Newell, Rev. Mod. Phys. {\bf 84}, 1527 (2012).
\end{thebibliography}
\end{document}